\documentclass[aps,pra,twocolumn,groupedaddress,amsmath,floatfix]{revtex4}
\usepackage{bm} 
\usepackage{graphicx}
\usepackage{epsfig}
\usepackage{subeqn}

\begin{document}


\title{Beyond mean field effects in the high momentum energy spectrum of a Bose gas: revisiting Beliaev's theory}

\author{Shai Ronen} 
\affiliation{JILA and Department of Physics, University of Colorado, Boulder, CO 80309-0440}
\email{sronen@colorado.edu}


\date{\today}

\begin{abstract}
The well-known Bogoliubov expression for the spectrum of a weakly
interacting dilute Bose gas becomes inadequate when the density or
interactions strength are increased. The corrections to the spectrum
due to stronger interactions were first considered by Beliaev
(S. T. Beliaev, Sov. Phys.-JETP, 7:289, (1958)). We revisit Beliaev's
theory and consider its application to a dilute gas with van der Waals
interactions, where the scattering length may be tuned via a Fano-Feshbach
resonance. We numerically evaluate Beliaev's expression for the
excitation spectrum in the intermediate momentum regime, and we also
examine the consequences of the momentum dependence of the two-body
scattering amplitude.These results are relevant to the interpretation
of a recent Bragg spectroscopy experiment of a strongly interacting
Bose gas.
\end{abstract}

\pacs{}

\maketitle

\section{INTRODUCTION \label{intro}}

A dilute Bose Einstein condensate with sufficiently weak interactions
is a many body system which can be described to a great degree of
success by mean field theory. On the other hand, condensed-matter
superfluids such as $^4$He are much more complex systems and difficult
to describe from first principles due to their high density and
stronger interactions, leading to, in superfluid Helium, a condensate
fraction of only about 15\% \cite{Woods73}. Superfluid Helium also
exhibits the interesting property of a roton minimum in its excitation
spectrum, which is absent in a weakly interacting dilute alkali
gas. It is therefore of theoretical and experimental interest to try
to bridge the two regimes by studying a Bose gas with stronger
interactions. 

Dilute gases with stronger interactions can be studied by considering
perturbation theory corrections to the mean field theory. Lee, Huang
and Yang (LHY), and Brueckner and Sawada
\cite{Lee57,Lee57a,Brueckner57}, first determined the quantum
depletion and the correction to the chemical potential of the ground
state of a homogeneous dilute Bose gas. It was found that the leading
corrections are proportional to the dimensionless parameter
$\sqrt{na^3}$ where $n$ is the density and $a$ the scattering length
of the atoms. Beliaev \cite{Beliaev58a,Beliaev58b} first applied the
methods of quantum field theory, specifically Green's functions, to
the description of a system of bosons. We shall distinguish between
the first order Beliaev's theory, which is equivalent to the
Bogoliubov spectrum for the case of contact interactions, and second
order theory, which gives $\sqrt{na^3}$ corrections to the excitation
spectrum. These two levels correspond roughly to first and second
order expansions of the Green's functions of the system in
perturbation theory. A description of the quantum field theory of
bosons and the Green's functions technique is available in textbooks
\cite{Abrikosov63,Fetter03}. Beliaev's work thus extends the LHY
results for the ground state to the excitation spectrum.

A dilute Bose gas with tunable interactions can be realized
experimentally with the utilization of a Fano-Feshbach resonance. A difficulty
arises in that the gas becomes more unstable with increasing
interaction strength due to 3-body collisions. Nevertheless, in the
shorter lifetime of a more strongly interacting gas, it is still
possible to observe relatively high frequency excitations (which
require short observation time, according to Heisenberg's uncertainty
principle). Such an experiment has been recently performed in JILA
\cite{Papp08b}.

Most studies of a dilute Bose gas assume an effective potential in the
form of contact interactions. This model is usually very successful,
but a correction is needed to describe excitations with higher
momenta, due to the momentum dependence of the scattering amplitude of
two colliding atoms. In section II we shall examine how this
correction can be incorporated in the first order theory of Beliaev
\cite{Beliaev58b}.

In section III we revisit Beliave's second order theory, now assuming,
for that purpose, a momentum independent scattering amplitude. Under
this condition, Beliaev derived a closed form integral expression for
the excitation spectrum. However, the integrals could only be
evaluated analytically in the low and high momentum regimes. It is
found that for low momenta (compared to $1/\xi$, where $\xi$ is the
healing length), the effect of stronger interactions is an upward
shift in the excitation frequency (compared to the Bogoliubov
spectrum), while at high momenta, the effect is to lower the
excitation frequency. These results were also re-derived using the
pseudopotential method \cite{Mohling60}. The interesting question that
arises is: what happens in the intermediate regime?  where is the
transition point from upward shift to downward shift?  This question
is of particular relevance to the experiment \cite{Papp08b} which
probed the spectrum at this intermediate regime. In this paper, we
answer this question by evaluating numerically the relevant
expressions. To the best of our knowledge (perhaps somewhat
surprisingly), this is the first time that Beliaev's theory's
prediction for intermediate momenta is examined.

\section{Effect of a momentum dependent scattering amplitude}

Beliaev's original first order theory\cite{Beliaev58b} already
incorporated, in general, the effect of a momentum dependent scattering
amplitude. Here, we shall take a few more steps to specialize it to
the description of a dilute gas with van der Waals interactions. We
shall begin with some definitions.We use units $\hbar=1$ and also let the mass of the bosons be $m=1$.

Let $n$ be the homogeneous total density of the gas, and $ n_0$ the
density of the condensed part. In the original paper, Beliaev
formulated his theory in terms of $n_0$. At the first order of theory,
which we discuss in this section, the two are equal. Quantum
depletion, giving rise to $n_0<n$, arise at the second order of the
theory. Let $\Psi_{\bm{p}}^{+}({\bm{r}})$ be a wave-function solution
of the Schr\"{o}dinger equation of the two-body scattering problem
with interaction potential $U(r)$, which behaves at infinity like a
plane wave $e^{i \bm{p} \cdot \bm{r}}$ and an outgoing spherical wave.
The scattering amplitude $\tilde{f}(\bm{p'},\bm{p})$ is related to the
$\Psi^{(+)}$-function by
\begin{eqnarray}
\tilde{f}(\bm{p'},\bm{p})=\int e^{-i \bm{p'}\cdot \bm{r}}U(\bm{r})\Psi^{(+)}_{\bm{p}}(\bm{r})d\bm{r}.
\label{eq:scatamp}
\end{eqnarray}
Notice that this definition differs by a numerical factor of $-4 \pi$
from the usual one. The usual scattering amplitude is the value of
$\tilde{f}(\bm{p'},\bm{p})$ at $|\bm{p}'|=|\bm{p}|$ ('on the energy shell').
Eq.~(\ref{eq:scatamp}) thus generalizes the usual scattering amplitude
to define an 'off the energy shell' amplitude.  We shall also need a symmetrized amplitude defined as:
\begin{eqnarray}
\tilde{f}_s(\bm{p'}, \bm{p}) \equiv \frac{ \tilde{f}(\bm{p'},\bm{p})+\tilde{f}(-\bm{p'}, \bm{p}) }{2}.
\end{eqnarray}

According to Beliaev's first order theory, the excitation spectrum of a dilute Bose gas is given
by the dispersion relation:
\begin{eqnarray}
\epsilon_{\bm{p}}^{(1)}=\sqrt{\left[\epsilon_{\bm{p}}^0+2 n_0
\tilde{f}_s(\frac{\bm{p}}{2},\frac{\bm{p}}{2})-n_0 \tilde{f}(0,0)
\right]^2-n_0^2|\tilde{f}(\bm{p},0)|^2},
\label{Bel1}
\end{eqnarray}
where $\epsilon_{\bm{p}}^0 \equiv p^2/2$ is the free particle kinetic
energy. The superscript $(1)$ in $\epsilon_{\bm{p}}^{(1)}$ indicates
first order. In the case of a gas with van der
Waals interactions, it is possible to derive analytic expressions for
the scattering amplitude.  At low energies, s-wave scattering is the
dominant process so that for on shell scattering $p=p'$ we may define
an isotropic amplitude
$\tilde{f}(p)=\tilde{f}(\bm{p'},\bm{p})$. From the usual
definition of the scattering length $a$ we have, $\tilde{f}(0) \equiv
\tilde{f}_0 = 4 \pi a$.

The scattering amplitude is related to the s-wave phase shift $\delta_0(p)$ as 
\begin{eqnarray}
\tilde{f}(p)=\frac{-4 \pi}{p}\frac{1}{\cot \delta_0(p)-i}
\end{eqnarray}
From the standard effective-range expansion, $p \cot \delta_0(p)=-\frac{1}{a}+\frac{1}{2}r_e p^2 +...,$ where $r_e$ is the
effective range. Thus
\begin{eqnarray}
\tilde{f}(p)=\frac{4 \pi a}{1-\frac{1}{2} a r_e p^2 + i a p}.
\label{eq:effrange}
\end{eqnarray}

For neutral atoms with van der Waals interactions, the following
useful relationship exists between the scattering length $a$, the
effective range $r_e$, and the $C_6$ coefficient \cite{Gao98,Fu03}:
\begin{eqnarray}
r_e/\beta_6=\left(\frac{2}{3 x_e}\right )\frac{1}{(a/\beta_6)^2}\left[{1+\left[ 1-x_e(a/\beta_6)\right]^2}\right],
\label{eq:gao}
\end{eqnarray}
where $\beta_6=(m C_6/\hbar^2)^{1/4}$ is a length scale associated
with the van der Waals interaction, and $x_e=[\Gamma(1/4)]^2/(2\pi)$,
with $\Gamma$ being the usual Gamma function. For clarity, we have written here
explicitly factors of $\hbar$ and $m$. Typical values of $\beta_6$ for alkali atoms are
about 100 bohr.

By extending the ideas of \cite{Gao98}, we find that the off-shell
amplitude $\tilde{f}(\bm{p},0)$ which appears in Eq.~(\ref{Bel1}), is
given, for a van der Waals interaction, by:
\begin{eqnarray}
\tilde{f}(\bm{p},0)=4\pi a (1+\frac{1}{2} b p^2+...),
\label{eq:offshell}
\end{eqnarray}
with 
\begin{eqnarray}
b=-\frac{1}{30} \beta_6^2 \sqrt{\pi} \left( -5 \sqrt{2} \beta_6/a+6 \frac{\Gamma(9/4)}{\Gamma(7/4)} \right).
\label{eq:offshellexp}
\end{eqnarray}
We have checked this analytic expression for the off shell amplitude
with numerical calculations using a model potential of $^{85}$Rb, with
good agreement.

A few notes are in place. First, the appearance of the factor $\frac{1}{2}$ in the
expression $\tilde{f}_s(\frac{\bm{p}}{2},\frac{\bm{p}}{2})$ is due to
moving to the center of mass system of an excited particle with
momentum $p$ and a condensate particle with zero momentum. Second, in
using Eq.~(\ref{eq:effrange}) with Eq.~(\ref{Bel1}) one should
note that the scattering amplitude has an imaginary part.  This gives
rise to an imaginary part of the energy describing correctly, for high
momentum, the decay of a particle excitation due to the scattering of
the particle off the rest of the condensate's atoms.  However at low
momenta (the quasi-particle phonon regime), the decay is actually suppressed, and is
only seen in the next level of theory (Beliaev's second order
theory). At first order of the theory it is sufficient to take the
real part of the scattering amplitude when using it to calculate the
spectrum. The off shell scattering amplitude $\tilde{f}(\bm{p},0)$ is
always real.

for $p a \ll 1$, eq.~(\ref{Bel1}) reduces to the familiar Bogoliubov spectrum:
\begin{eqnarray}
\epsilon_{\bm{p}}^{(1)}=\sqrt{ (\epsilon_{\bm{p}}^0)^2+2 n_0 \tilde{f}_0 \epsilon_{\bm{p}}^0}.
\label{eq:BdG}
\end{eqnarray}
By increasing $p a$ one enters the regime where finite-momentum
corrections are necessary. For a fixed $p$, this could happen due to
increasing $a$. If the scattering length is larger than the van der
Waals length scale, i.e, $a \gg \beta_6$, then according to
Eq.~(\ref{eq:gao}), $r_e \approx 1.4 \beta_6$, while according to
Eq.~(\ref{eq:offshellexp}), $b \approx 0.44 \beta_6^2$. If the
condition $p \beta_6 \ll 1$ is still satisfied (which is normally the
case), then the correction to the off shell amplitude
$\tilde{f}(\bm{p},0)$ is very small and one can take it to be equal to
$a$. Furthermore, in Eq.~(\ref{eq:effrange}), the effective range
contribution due to non-zero $r_e$ is smaller by a factor $p \beta_6$
than the universal universal factor $i p a$. Therefore, we arrive in a
final universal momentum dependent dispersion relation, valid
for $p \beta_6 \ll 1$:
\begin{eqnarray}
\epsilon_{\bm{p}}^{(1)}=\sqrt{\left[\epsilon_{\bm{p}}^0+ \frac{8 \pi
n_0 a}{1+ (pa)^2/4}-4 \pi n_0 a \right]^2-n_0^2 (4 \pi a)^2}.
\label{eq:univ}
\end{eqnarray}

\begin{figure}
\resizebox{3.5in}{!}{\includegraphics{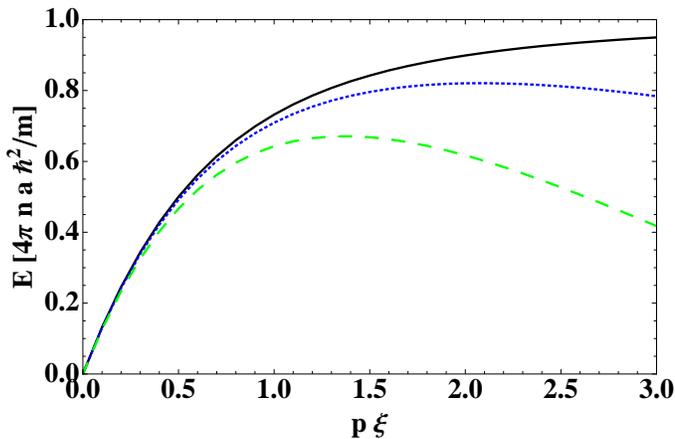}}
\caption{
\label{fig:ka}
Change in line shift due to the momentum dependence of the scattering
amplitude. The three curves correspond to different values of the
parameter $a/\xi=\sqrt{8 \pi n a^3}$. The values are 0 (solid black),
0.2 (dotted blue), and 0.4 (dashed green)}
\end{figure}

At this level of the theory, we may replace $n_0$ in
Eq.(\ref{eq:univ}) with $n$. We define an energy shift
$E=\epsilon_{\bm{p}}^{(1)}-\epsilon_{\bm{p}}^0$ which is the difference between
the excitation energy of a quasi-particle with momentum $p$ in the
condensate, and the kinetic energy of a free particle with the same
momentum.  In Fig.~(\ref{fig:ka}) we plot the energy shift as a
function of the dimensionless parameter $p \xi$ where
$\xi=\frac{1}{\sqrt{8 \pi n a}}$ is the healing length. The shift is
shown for different values of the parameter $a/\xi=\sqrt{8 \pi na^3}$.
The curve corresponding to $a/\xi=0$ should be interpreted as taking
the limit of $\sqrt{n a^3} \rightarrow 0$ (notice also that the
vertical axis is proportional to $n a$), and gives the usual
Bogoliubov spectrum, with the asymptotic shift value of $4 \pi n a$
for large momentum. Increasing $a/ \xi$ gives rise to observable
deviations at high momentum. At small momenta the shift is linear in
momentum and the slope gives the speed of sound in the
condensate. Since the change in the real part of the scattering
amplitude is proportional to $p^2$, the slope of all curves is the
same at small momenta, and the speed of sound is unchanged by this
effect.

\section{Effect of quantum depletion}

The second order of Beliaev's theory is associated with the LHY
quantum depletion, but is concerned mainly with the excitation
spectrum rather than the ground state, and includes other second order
processes such as Beliaev's damping (the decay of a quasi-particle
into two lower energy quasi-particles). Beliaev summed the relevant
Feynman diagrams which contribute to the Green's function of the
system up to second order. The excitation spectrum is then found by
the poles of the Green's function in the $(\bm{p},p^0)$ space, with
$\bm{p}$ momentum and $p^0$ energy. Mohling and Sirlin
\cite{Mohling60} arrived at the same excitation spectrum using a more
straightforward application of the LHY pseudopotential method. We
shall briefly state here the results in the language of Green's
functions \cite{Fetter03,Abrikosov63}. In this section we use the
4-momentum notation $p \equiv (\bm{p};p^0)$. In a Bose condensate there are
two Green's functions, a regular Green's function $G_{11}(p)$, and an
anomalous Green's function $G_{12}(p)$. Correspondingly, there are two
self-energies, a regular self-energy $\Sigma_{11}(p)$ and an anomalous
self-energy $\Sigma_{12}(p)$. Also, we have the free particle Green's
function $G^0(p)$.  The Dyson's equations of this system can written
in the matrix form:

\begin{eqnarray}
\bm{G}(p)=\bm{G}^{0}(p)+\bm{G}^0(p)\bm{\Sigma}(p)\bm{G}(p),
\label{eq:Dyson}
\end{eqnarray}
with
\begin{eqnarray}
\bm{G}(p)=\left[ \begin{array}{cc}
G_{11}(p) & G_{12}(p) \\
G_{12}(-p) & G_{11}(-p)  \end{array} \right],
\end{eqnarray}
\begin{eqnarray}
\bm{G}^0(p)=\left[ \begin{array}{cc}
G^0(p) & 0 \\
0 & G^0(-p) \end{array} \right],
\end{eqnarray}
\begin{eqnarray}
\bm{\Sigma}(p)=\left[ \begin{array}{cc}
\Sigma_{11}(p) & \Sigma_{12}(p) \\
\Sigma_{12}(-p) & \Sigma_{11}(-p) \end{array} \right].
\end{eqnarray}

The self-energies can be calculated in perturbation theory using
Feynman's diagrams. Then, the Green's functions can be obtained by
solving the algebraic matrix equation~(\ref{eq:Dyson}), and their poles give the excitation
spectrum (both kinds of Green's functions have the same
poles). Following the original papers, we assume in this section
(in contrast to the previous one) a \textit{momentum independent}
scattering amplitude, valid for $p a \ll 1$, i.e,  $\tilde{f}_0=4\pi a$.  The first order of theory gives:

\begin{eqnarray}
\Sigma_{11}(p)^{(1)}=2 n_0 f_0 \\
\Sigma_{12}(p)^{(1)}=n_0 f_0. 
\end{eqnarray}
When these expressions are used to calculate the excitation spectrum
via the poles of the corresponding Green's functions, one retrieves
the ordinary Bogoliubov spectrum, Eq.~(\ref{eq:BdG}).

The second order contributions (additive to the first order), written
$\Sigma_{11}^{(2)}$ and $\Sigma_{12}^{(2)}$, are:
\begin{widetext}
\begin{eqnarray}
\label{eq:self}
\Sigma_{12}^{(2)}(p) & = &  \nonumber \frac{1}{2} n_0 \tilde{f}_0^2
 \frac{1}{(2 \pi)^3} \int \frac{d
 \bm{q}}{\epsilon_{\bm{q}}^{(1)}\epsilon_{\bm{k}}^{(1)}} R(\bm{q},\bm{k}) \times \\ \nonumber
 & & \left[ \frac{1}{p^0 -
 \epsilon_{\bm{q}}^{(1)}-\epsilon_{\bm{k}}^{(1)}+i\delta}-
 \frac{1}{p^0 +
 \epsilon_{\bm{q}}^{(1)}+\epsilon_{\bm{k}}^{(1)}-i\delta} \right]+ \\ \nonumber 
 & & \frac{1}{\pi^2}\sqrt{n_0} \tilde{f}_0^3 n_0 \tilde{f}_0, \\ 
 \Sigma_{11}^{(2)}(p) & = &  \frac{1}{2} n_0 \tilde{f}_0^2
 \frac{1}{(2\pi)^3}\int \frac{d\bm{q}}{\epsilon_{\bm{q}}^{(1)}
 \epsilon_{\bm{k}}^{(1)}} \times \\ \nonumber 
 & & \left[
 \frac{Q^{-}(\bm{q},\bm{k})}{p^0-\epsilon_{\bm{q}}^{(1)}-\epsilon_{\bm{k}}^{(1)}+i\delta}-
 \frac{Q^{+}(\bm{q}.\bm{k})}{p^0+\epsilon_{\bm{q}}^{(1)}+\epsilon_{\bm{k}}^{(1)}-i\delta}+
 \epsilon_{\bm{q}}^{(1)}+\epsilon_{\bm{k}}^{(1)} \right] + \\ \nonumber 
 & & \frac{8}{3\pi^2}\sqrt{n_0 f_0^3} n_0 f_0,
\end{eqnarray}
where $\bm{k} \equiv \bm{p}-\bm{q}$, and the functions $R, Q^{\mp}$ are:
\begin{eqnarray}
R(\bm{q},\bm{k}) & = & 2 \epsilon^0_{\bm{q}}
\epsilon^0_{\bm{k}}-2\epsilon_{\bm{q}}^{(1)}
\epsilon_{\bm{k}}^{(1)}+n_0^2 f_0^2, \\ \nonumber
Q^{\mp}(\bm{q},\bm{k}) & = & 3 \epsilon^0_{\bm{q}}
\epsilon^0_{\bm{k}}+n_0 f_0
(\epsilon_{\bm{q}}^0+\epsilon_{\bm{k}}^0)+n_0^2 f_0^2 \mp \left[ n_0
f_0
(\epsilon_{\bm{q}}^{(1)}+\epsilon_{\bm{k}}^{(1)}-\epsilon_{\bm{q}}^{(1)}\epsilon_{\bm{k}}^0-\epsilon_{\bm{k}}^{(1)}
\epsilon_{\bm{q}}^0 \right].
\end{eqnarray}
\end{widetext}
The $i \delta$ factors in Eq.~(\ref{eq:self}) are needed for
convergence and should be understood in the usual sense of the limit
$\delta \rightarrow 0^{+}$. By solving the Dyson's matrix equation
with these self energies, the Green function $G_{11}$ can be expressed
in the convenient form (showing its poles):
\begin{eqnarray}
G_{11}(p)=\frac{A_p}{p^0-\epsilon_{\bm{p}}^{(1)}-\Lambda^{-}(p)}-\frac{B_p}{p^0+\epsilon_{\bm{p}}^{(1)}+\Lambda^{+}(p)},
\end{eqnarray} 
where $A_p$ and $B_p$ are certain functions of $p$ (with no poles),
and $\Lambda^{\mp}(p)$ are second-order corrections
\begin{eqnarray}
\Lambda^{\mp}(p) & = & \frac{\epsilon_p^0}{2\epsilon_{\bm{p}}^{(1)}}
(\Sigma_{11}(p) +\Sigma_{11}(-p)-2\mu)^{(2)}+ \\ \nonumber
&& +\frac{n_0 f_0}{2\epsilon_{\bm{p}}^{(1)}}(\Sigma_{11}(p)+  
\Sigma_{11}(-p)-2\mu-2\Sigma_{12}(p))^{(2)} \pm \\ \nonumber
&& \pm \frac{1}{2}(\Sigma_{11}(p)-\Sigma_{11}(-p))^{(2)},
\end{eqnarray}
with  $\mu^{(2)}$ the second order correction to the chemical potential,
\begin{eqnarray}
\mu^{(2)}=(5/3\pi^2)\sqrt{n_0 \tilde{f}_0^3} n_0 \tilde{f}_0.
\end{eqnarray}

Inspecting the poles of the Green's function, the excitation spectrum in this approximation  is then given by:
\begin{eqnarray}
\epsilon_{\bm{p}}=\epsilon_{\bm{p}}^{(1)}+\Lambda^{-}(\bm{p};\epsilon_{\bm{p}}^{(1)}).
\label{eq:spectrum}
\end{eqnarray}

In general, there is no known analytic solution to the integrals in
Eq.~(\ref{eq:self}). Beliaev derived an analytic solution in the the
limit of small momenta $p \xi \ll 1$: 
\begin{eqnarray}
\label{eq:lowp}
\epsilon_{\bm{p}} & = & p \sqrt{n \tilde{f}_0} \left(
1+\frac{1}{\pi^2} \sqrt{n \tilde{f}_0^3} \right) - \\ \nonumber & - &
i \frac{3 p^5}{640 \pi n} \; \; \; (p \xi \ll 1).
\end{eqnarray}
This result is expressed here in terms of the total density $n$ rather
than the condensate density $n_0$. The relation between $n$ and $n_0$
is:
\begin{eqnarray}
n-n_0= n_0 \sqrt{n_0 \tilde{f}_0^3} /(3 \pi^2).
\end{eqnarray}
According to Eq.~(\ref{eq:lowp}), the second order effects give rise
to a upward shift in the excitation energy with linear dependence in
momentum, and thus a shift in the velocity of sound:
\begin{eqnarray}
c=\sqrt{\tilde{f}_0 n}\left[1+\frac{1}{\pi^2}(n \tilde{f}_0^3)^{\frac{1}{2}}\right].
\end{eqnarray}
This result for the speed of sound, derived above from the slope of the one-particle
excitation spectrum at low momentum, is identical to that obtained by considering only the LHY expression
for the ground state energy:
\begin{eqnarray}
\frac{E}{V}=\frac{ n^2 \hbar^2
\tilde{f}_0}{2m}\left[1+\frac{16}{15 \pi^2} (n \tilde{f}_0^3)^{\frac{1}{2}}\right],
\end{eqnarray} 
where we explicitly included factors of $m$ and $\hbar$, and using the
thermodynamic relations $P=-\left(\frac{\partial E}{\partial V}
\right)_N$ for the pressure $P$, and $m c^2= \frac{\partial
P}{\partial n}$ \cite{Fetter03}.  Thus, from the macroscopic
thermodynamic point of view, the increase in speed of sound is
traced to the equation of state becoming stiffer due to quantum
depletion. Eq.~(\ref{eq:lowp}) also has an imaginary part which is due to the decay of one
quasi-particle into two other quasi-particles. 

On the other hand, Mohling and Sirlin \cite{Mohling60} give an
expression for the shift at high momenta, with $p \xi \gg 1$, (but, at
the same time, $p a \ll 1$). The result may be written in the form:
\begin{eqnarray}
\label{eq:highp}
\epsilon_{\bm{p}} & = & \epsilon_{\bm{p}}^0+ 2 n \tilde{f}_0-\mu_{LHY} - \\ \nonumber
& - & i n \sigma  \frac{|\bm{p}|}{2} \; \; \; \; (p \xi \gg 1),
\end{eqnarray}
with the LHY chemical potential :
\begin{eqnarray}
\mu_{LHY}=n \tilde{f_0} \left( 1+\frac{4}{3 \pi^2}\sqrt{n \tilde{f}_0^3} \right),
\end{eqnarray}
and the cross section for two identical bosons
\begin{eqnarray}
\sigma=8 \pi a^2.
\end{eqnarray}
We may understand the result of Eq.~(\ref{eq:highp}) as follows: the
excitation energy is the difference between final state and initial
state energies. The initial state energy of the (quasi-)particle is
$\mu$, and the final state energy at high momentum is the free kinetic
energy $\epsilon_{\bm{p}}^0$ plus the sum of direct and exchange
interactions of the excited particle with all other, low momentum
particles, $2 n_0 \tilde{f}_0$. In addition, the excitation decays due
to collision of the high momentum particle with all other particles,
with cross section $\sigma$ and center of mass momentum $\bm{p}/2$.
Note, that Eq.~(\ref{eq:highp}) only considers the high momentum ($p
\xi \gg 1$) effect due the many body interactions. The correction due
to the momentum-dependent scattering amplitude, as described by
Eq.~(\ref{eq:univ}) can be neglected here as long as $p a\ll 1$. The
required conditions, $1/\xi \ll p \ll 1/a$, can in principle be
satisfied simultaneously for a sufficiently dilute gas.

What happens in the intermediate range, $p \xi \approx 1$ ? We have
calculated numerically the excitation energy
Eq.~({\ref{eq:spectrum}). This was done by deforming the path of
integration from the real line into the complex plane around the
singularities indicated by the $i \delta$ factors.  The result is
shown in Fig.~(\ref{fig:dep}). The plot shows the quantity
$\Re(\epsilon_{\bm{p}}-\epsilon_{\bm{p}}^{(1)})$, i.e the difference
between the real part of Beliaev's excitation energy and the
Bogoliubov energy. At low momenta there is an upward linear shift
consistent with Eq.~(\ref{eq:lowp}). This linear shift is consistent
with the correction to the speed of sound derived by LHY theory. The
linear behavior holds up to $p \xi \approx 0.6$. At high momentum
there is a down shift with very slow $~1/p$ convergence towards
the asymptotic value $\frac{4}{3}\tilde{f}_0 n\sqrt{\tilde{f}_0^3
n}/\pi^2$ consistent with Eq.~(\ref{eq:highp}). In between, we find
the transition point at $p \approx 6.18 \xi^{-1}$, where the second
order contribution is zero, so that at this point the excitation
energy to second order is identical with the first order Bogoliubov's
theory.
\begin{figure}
\resizebox{3.5in}{!}{\includegraphics{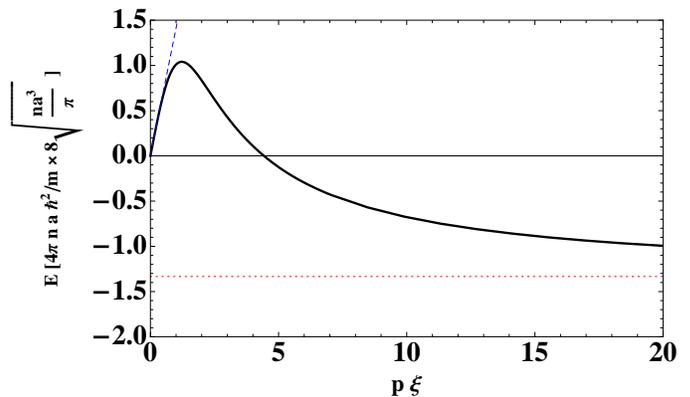}}
\caption{
\label{fig:dep}
Solid black line: change in line shift due to quantum depletion and other second-order
processes. The correction is additive to the solid black line in
Fig.~(\ref{fig:ka}). The amount of correction is proportional to $8\sqrt{n a^3/
\pi}$, as indicated on the vertical axis. Blue dashed line: the linear asymptote at small momenta
which is consistent with LHY theory. Red dotted line: the asymptotic value at high momenta.}
\end{figure}

The imaginary part of the excitation energy gives rise to
quasi-particle damping with an effective cross-section
$\sigma_{\bm{p}}=-2 \Im(\frac{\epsilon_{\bm{p}}}{p n})$ ranging from
$0$ at $p=0$ to $ 8\pi a^2$ at $p \xi \gg 1$. It is shown in
Fig.~(\ref{fig:decay}).  This damping cross-section has been
calculated before using another method and measured in experiment
\cite{Katz02}. The two methods of calculation give identical results.

\begin{figure}
\resizebox{3.5in}{!}{\includegraphics{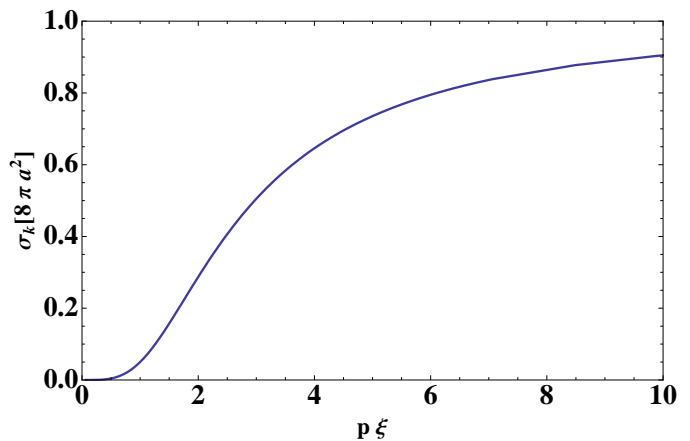}}
\caption{
\label{fig:decay}
Cross section for quasi-particle decay.}
\end{figure}

\subsection{Discussion and Conclusions}

In this paper we have revisited Beliaev's theory with emphasis on two
mechanisms for a shift in the excitation spectrum of a Bose gas from
the usual Bogoliubov theory. One is due to the momentum dependence of
scattering amplitude, when the condition $k a \ll 1$ is not valid,
which generally causes a downward shift in the excitation frequency at
higher momentum, as demonstrated in Fig.~(\ref{fig:ka}). The other is a
purely many body effect, occurring when the condition $\sqrt{8 \pi n
a^3} \ll 1$ is not obeyed. This effect is associated with quantum
depletion due to interactions as well as with quasi-particle decay. As
shown in Fig.~(\ref{fig:dep}). It gives rise to upward shift at low
momentum and downward shift at high momentum. 

We note that the two separate effects were calculated
under exclusive conditions.  The effect due to momentum dependent
scattering amplitude ($p a \approx 1$) was calculated in first order of the
interactions, valid for $\sqrt{8 \pi na^3} \ll 1$ . On the other, the
second order effects describing $\sqrt{8 \pi n a^3}$ corrections
assumed $p a \ll 1$. Currently, there is no known result that holds in
the regime where both the momentum dependence of the scattering
amplitude and quantum depletion due to strong interactions are
important. The physical interpretation that follows Eq.~(\ref{eq:highp}) suggests that
at high enough momentum the shift due to these two effects should be additive and combine to give
\begin{eqnarray}
\label{eq:asym}
\epsilon_{\bm{p}} & = & \epsilon_{\bm{p}}^0+ 2 n \Re\left[\tilde{f}_s(p/2)\right]-\mu_{LHY} - \\ \nonumber
& - & i n \sigma  \frac{|\bm{p}|}{2} \; \; \; \; (p \xi \gg 1).
\end{eqnarray}
This expression assumes the asymptotic value the of the energy shown
in of Fig.~(\ref{fig:dep}). Because of the slow convergence, even at
$p \xi = 20$ the energy is still only 75\% of the asymptotic value.
The Bragg spectroscopy experiment in Ref.~\cite{Papp08b} explored the
regime where both effects discussed above are important, and showed
evidence for a downward shift in the excitation energy of a strongly
interacting gas. But for the measurements with the strongest
interactions, which showed the largest shifts, the experiment had $p
\xi \approx 2$, where Fig.~(\ref{fig:dep}) shows upward rather than
downward shift. On the other hand, the prediction of down shift due to
the 2-body momentum dependence of the scattering length seems to go
some way towards explaining the observations.

Finally, we note that the dynamic structure factor is a quantity more directly
related to the actual measurement process of a Bragg experiment
\cite{Steinhauer02, BEC2003}. A partial attempt at incorporating
second order effects due to strong interactions in the dynamic
structure factor is found in Ref.~\cite{Mohling65}. It would be of
interest to complete this program.

\begin{acknowledgments}
I thank John Bohn, Debbie
Jin, Scott Papp, Eric Cornell, Victor Gurarie and Jami Kinnunen for helpful discussions. I acknowledge financial support
from the NSF.
\end{acknowledgments}

\bibliography{biblo}

\end{document}